%% file: main.tex
\PassOptionsToPackage{table}{xcolor} 
\documentclass[11pt]{article}

\usepackage[preprint]{acl}

\usepackage{times}
\usepackage{latexsym}

\usepackage[T1]{fontenc}

\usepackage[utf8]{inputenc}

\usepackage{microtype}

\usepackage{inconsolata}
\usepackage{subcaption}

\usepackage{graphicx}
\usepackage{microtype}      
\usepackage{hyperref}       
\usepackage{url}            
\usepackage{amsmath}        
\usepackage{amssymb}        
\usepackage{enumitem}       
\usepackage{xspace}         
\usepackage{arydshln}

\usepackage{booktabs}       
\usepackage{array}          
\usepackage{tabularx}       
\usepackage{longtable}      
\usepackage{siunitx}        

\usepackage{tcolorbox}      
\usepackage{xcolor}  
\usepackage{multirow}       
\usepackage{wrapfig}        
\usepackage{lineno}         
\usepackage{lipsum}         
\usepackage{nicematrix}     
\usepackage{soul}
\usepackage{adjustbox}
\usepackage{cleveref}

\title{\ourdataset: Advancing Dual-Mode Multilingual Distillation for Software Engineering Agents}

\author{Wasi Uddin Ahmad, Nikolai Ludwig, Somshubra Majumdar, Boris Ginsburg \\ 
NVIDIA \\ 
\texttt{\{wasiuddina,nliudvig\}@nvidia.com} \\ 
\texttt{\url{https://huggingface.co/collections/nvidia/open-swe-traces}}
}

\definecolor{darkblue}{rgb}{0, 0, 0.5}
\hypersetup{colorlinks=true, citecolor=darkblue, linkcolor=darkblue, urlcolor=darkblue}

\usepackage{listings}

\definecolor{dark-gray}{gray}{0.85}
\definecolor{light-gray}{gray}{0.95}
\definecolor{mygreen}{rgb}{0,0.4,0}
\definecolor{mygray}{rgb}{0.5,0.5,0.5}
\definecolor{mymauve}{rgb}{0.58,0,0.82}
\definecolor{myred}{rgb}{0.82, 0.1, 0.26}

\usepackage[T1]{fontenc}
\usepackage{textcomp} 
\usepackage{xcolor}

\lstdefinestyle{CustomPy}{
    language=Python,
    upquote=true,               
    showstringspaces=false,
    basicstyle=\small\ttfamily,
    keywordstyle=\bfseries\color{green!40!black},
    commentstyle=\itshape\color{purple},
    stringstyle=\color{blue},   
    numbers=left,
    numberstyle=\tiny\color{gray},
    breaklines=true,
    columns=flexible,           
    escapeinside={(*@}{@*)}     
}

\usepackage{textcomp}

\newcommand{\ourdataset}{\textsc{Open-SWE-Traces}\xspace}
\newcommand{\ourmodel}{\textsc{Open-SWE-Agent}\xspace} 
\newcommand{\ourthinkmodel}{\textsc{Open-SWE-Agent-Thinking}\xspace}
\newcommand{\ournonthinkmodel}{\textsc{Open-SWE-Agent-Instruct}\xspace}

\usepackage{fvextra}

\begin{document}

\setlength{\abovedisplayskip}{2pt}
\setlength{\belowdisplayskip}{2pt}

\maketitle
\begin{abstract}
The path toward autonomous software engineering is currently bottlenecked by a severe deficit of diverse, large-scale trajectory data. We address this by introducing \ourdataset, an expansive dataset of 207,489 agentic trajectories spanning nine programming languages (Python, Go, TS, JS, Rust, Java, PHP, C, C++). Sourced from 20,000 real-world PRs via OpenHands and SWE-agent harnesses, the dataset utilizes a hybrid-reasoning synthesis: Minimax-M2.5 generates trajectories with explicit "thinking" processes, while Qwen3.5-122B provides high-quality "non-thinking" traces.
Filtered for permissive licenses (MIT, Apache, BSD) from SWE-rebench-V2, this data facilitates the training of models capable of long-horizon reasoning. We validate the dataset by fine-tuning the Qwen3-30B-A3B series (Thinking, Coder, and Instruct). The best performing model achieves resolve rates of {\bf 61.7}\% on SWE-bench Verified, {\bf 57.1}\% on SWE-bench Multilingual, and {\bf 36.8}\% on SWE-bench Pro. These results establish \ourdataset as a premier resource for distilling human-level software engineering capabilities into efficient, open-source agentic LLMs.
\end{abstract}

\section{Introduction}
\label{sec:introduction}

\input{sections/1_introduction}

\section{\ourdataset: From Trajectory Synthesis to Quality Filtering}
\label{sec:data}

\input{sections/2_dataset}

\section{Experiments}
\label{sec:experiments}

\input{sections/3_experiments}

\section{Ablation and Analyses}
\label{sec:ablation_and_analysis}

\input{sections/4_analysis}

\section{Related Works}
\label{sec:related_works}

\input{sections/5_related_works}

\section{Conclusion}
\label{sec:conclusion}
\input{sections/6_conclusion}


\section*{Limitations}
While our distillation framework achieves state-of-the-art results, it is subject to several constraints. First, the performance and behavioral profile of our agents are inherently tethered to the teacher models (Minimax-M2.5 and Qwen3.5-122B); consequently, any latent biases or systemic errors present in these predecessors are likely inherited by the students. Furthermore, the inherent stochasticity of LLMs, compounded by the volatility of software execution environments, introduces unavoidable variance. Although we mitigated this through triple-run aggregation, infrastructure dependencies and environmental factors can cause minor fluctuations in scores, which may affect precise reproducibility.

\section*{Ethics Statement}
We are committed to the public release of \ourdataset and have implemented rigorous protocols to ensure its integrity. All agent trajectories were derived from open-source repositories under permissive licenses (e.g., MIT, Apache 2.0, BSD), with automated filtering pipelines employed to redact Personally Identifiable Information (PII) and sensitive credentials. 
Furthermore, we utilized Gemini to assist in refining the manuscript’s linguistic clarity and to generate the visual data flow illustration (Figure \ref{fig:dataset_creation}). Nonetheless, the authors have meticulously reviewed all suggestions and remain fully responsible for the research findings, technical accuracy, and final content.

\bibliography{bib/custom,bib/anthology-1,bib/anthology-2}

\input{sections/appendix}

\end{document}

%% file: sections/1_introduction.tex
The landscape of software engineering (SWE) has been fundamentally reshaped by the emergence of Large Language Model (LLM)-driven agents \citep{cao2026qwen3, kimiteam2026kimik25visualagentic, minimax2026m25, liu2025deepseek, google_gemini3pro_2025, openai2025gpt52, anthropic2025b}. These systems do not merely suggest code snippets; they navigate complex repositories, interpret tool feedback, and autonomously iterate on patches. The industry standard for assessing these agents is repository-level issue resolution, popularized by benchmarks such as SWE-bench \citep{jimenez2023swe, chowdhury2024introducing, deng2025swe}, where success is measured by an agent’s ability to resolve real-world bugs verified against a project’s internal test suite.

As the field expands toward multilingualism with benchmarks like Multi-SWEbench \citep{zan2026multi} and SWE-PolyBench \citep{rashid2025swe}, a critical gap has emerged between evaluation and development. 
While evaluation datasets have become increasingly diverse, the community lacks the large-scale interaction traces and pre-built environments necessary for robust model development.
This resource scarcity changed with the release of SWE-Rebench v2 \citep{badertdinov2026swe}, which introduced a massive collection of over 32,000 containerized SWE tasks spanning 20 programming languages. Utilizing this unprecedented scale of executable environments, we bridge the gap between static evaluation and large-scale agent training by releasing \ourdataset: a comprehensive corpus of 207,489 synthesized trajectories across nine programming languages.

\input{tables/data_stat_2}

\ourdataset is uniquely designed to facilitate Dual-Mode Multilingual Distillation, a strategy inspired by recent flagship foundation models such as Qwen3.5 \citep{qwen3.5} and Qwen3.6 \citep{qwen3.6-27b}. These systems have pioneered a unified "switchable" reasoning framework, integrating both a thinking mode for complex, multi-step planning and a non-thinking mode for rapid, execution-oriented responses. To support this, we employ an ensemble of state-of-the-art models—specifically MiniMax-M2.5 \citep{minimax2026m25} for reasoning-heavy traces and Qwen3.5-122B \citep{qwen3.5} for diverse tool-use—to capture both high-fidelity reasoning traces and direct behavioral logs.

The necessity for this distinction is empirically supported by our trajectory statistics, as shown in \cref{tab:data_stat_2}. We observe that trajectories incorporating internal "thinking" (Chain-of-Thought) demonstrate a remarkable capacity for trajectory compression. Within the OpenHands framework, thinking-enabled traces reduce the average assistant turns per trajectory from 94.08 to 58.22—a 38\% increase in execution efficiency. While these thinking steps increase the average tokens per turn, they effectively prune unproductive trial-and-error loops. \ourdataset allows models to reserve intensive reasoning for difficult problems while maintaining execution speed for standard operations.
To ensure technical integrity, our trajectory generation pipeline incorporates rigorous multi-stage quality filtering and a validation process. This framework uses AST-based auditing to eliminate "git hacking" behaviors, where agents attempt to bypass authentic problem-solving by extracting solutions from repository metadata.

The primary contribution of this work is the release of \ourdataset as a foundational resource for distilling human-level software engineering capabilities into efficient, open-source agentic models. We validate this by fine-tuning the Qwen3-30B-A3B series; our best performing model achieves state-of-the-art resolve rates: {\bf 61.7}\% on SWE-bench Verified, {\bf 57.1}\% on SWE-bench Multilingual, and {\bf 36.8}\% on SWE-bench Pro. These results demonstrate that \ourdataset effectively enables the training of the next generation of dual-mode autonomous agents.

Our contributions are summarized as follows:
\begin{itemize}[noitemsep, topsep=2pt]
    \item {\bf \ourdataset}: We introduce the largest multilingual trajectory corpus to date, featuring 207,489 high-fidelity agent traces. The dataset spans diverse agent harnesses and is uniquely structured with both "thinking" and "non-thinking" trajectories to support dual-mode agent development.
    \item {\bf Performant Dual-Mode Distillation}: We release \ourmodel, fine-tuned from Qwen3-Coder-30B-A3B using \ourdataset. Our agent demonstrates the efficacy of dual-mode training by achieving a {\bf 61.7}\% resolve rate on SWE-bench Verified, while maintaining strong performance on SWE-bench Multilingual ({\bf 57.1}\%) and SWE-bench Pro ({\bf 36.8}\%) benchmarks. These results establish \ourmodel as a highly capable open-source baseline for autonomous software engineering.
    \item {\bf Systematic Analysis}: We provide an extensive empirical study isolating the drivers of agentic performance. Our analysis evaluates the impact of base model selection, data filtering strategies (resolved-only vs. all), and the scaling effects of multilingual vs. Python-only distillation. Finally, we analyze the trade-offs between thinking and non-thinking modalities and demonstrate the model's generalization to unseen execution harnesses.
\end{itemize}

%% file: tables/data_stat_2.tex
\begin{table*}[ht]
\centering

\resizebox{\textwidth}{!}{%
\begin{tabular}{@{}lrrrr@{}}
\toprule
\textbf{Metric} & \textbf{\begin{tabular}[c]{@{}c@{}}OpenHands\\ (w/o thinking)\end{tabular}} & \textbf{\begin{tabular}[c]{@{}c@{}}OpenHands\\ (w/ thinking)\end{tabular}} & \textbf{\begin{tabular}[c]{@{}c@{}}SWE-Agent\\ (w/o thinking)\end{tabular}} & \textbf{\begin{tabular}[c]{@{}c@{}}SWE-Agent\\ (w/ thinking)\end{tabular}} \\ \midrule
Total Assistant Tokens & 2,078,918,832 & 1,106,238,984 & 1,803,168,804 & 1,816,518,905 \\
Total Assistant Turns & 7,225,661 & 3,529,736 & 7,705,056 & 6,788,458 \\
Avg. Assistant Turns per Traj. & 94.08 & 58.22 & 130.70 & 74.82 \\
Avg. Tokens per Assistant Turn & 287.7 & 313.4 & 234.0 & 267.6 \\ \bottomrule
\end{tabular}%
}
\vspace{-2mm}
\caption{Token and turn statistics for \ourdataset.}
\label{tab:data_stat_2}
\end{table*}

%% file: sections/2_dataset.tex
\begin{figure*}[htbp]
    \centering
    \includegraphics[width=1.0\textwidth]{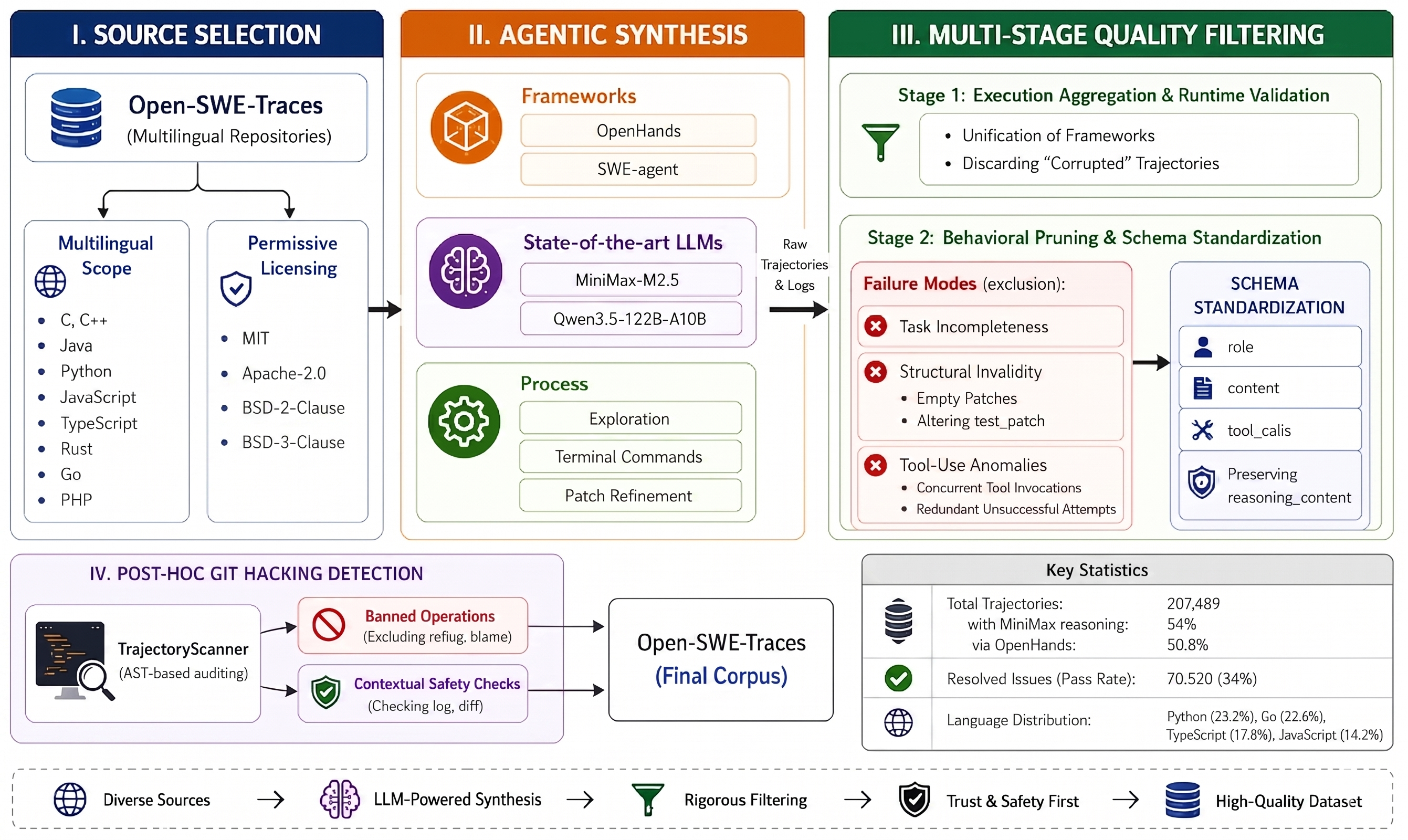}
    \caption{The \ourdataset Dataset Construction Pipeline.}
    \label{fig:dataset_creation}
\end{figure*}

The construction of \ourdataset follows a systematic pipeline designed to generate high-fidelity, multi-step trajectories across a diverse, multilingual landscape. As illustrated in \cref{fig:dataset_creation}, this workflow transitions from rigorous source selection to large-scale agentic synthesis, culminating in a multi-stage refinement process to retain only technically sound trajectories for distillation.

\subsection{Repository Selection and Criteria}
We anchor our trajectory generation in SWE-rebench v2 \citep{badertdinov2026swe}, a multilingual iteration of SWE-rebench \citep{badertdinov2025swe}. To ensure both the technical utility and legal integrity of the corpus, we applied the following strict selection heuristics:

\begin{itemize}[noitemsep, topsep=2pt]
    \item {\bf Multilingual Scope:} We limited instances to nine languages: Python, Java, C, C++, JavaScript, TypeScript, Rust, Go, and PHP.
    \item {\bf Permissive Licensing:} To remain compliant with redistribution standards, we enforced a strict licensing filter. Only repositories distributed under MIT, Apache-2.0, BSD (2-Clause and 3-Clause) licenses were included.
\end{itemize}

\subsection{Multilingual Trajectory Synthesis}
To simulate the complexities of real-world software engineering, we employed an ensemble of state-of-the-art LLMs to serve as the cognitive core for autonomous coding agents. Specifically, we integrated MiniMax-M2.5 \citep{minimax2026m25} and Qwen3.5-122B-A10B \citep{qwen3.5} into the OpenHands \citep{wang2024openhands} and SWE-agent \citep{yang2024swe} frameworks.

Within these environments, the models were tasked with navigating an agent-computer interface to explore unfamiliar codebases, execute terminal commands, and iteratively refine patches for reported issues. By deploying a heterogeneous set of model architectures, we aimed to capture a wider variety of reasoning heuristics and tool-use strategies. This diversity is critical for ensuring that the resulting trajectories represent a robust spectrum of problem-solving approaches rather than the idiosyncratic biases of a single model family.

\subsection{Multi-Stage Quality Filtering}
To refine raw execution logs into a high-fidelity corpus suitable for model distillation, we implemented a rigorous two-stage filtering and normalization pipeline. This process ensures that only trajectories demonstrating coherent problem-solving and technical correctness are preserved.

\paragraph{Stage 1: Execution Aggregation and Runtime Validation}
The initial stage focuses on the structural and functional integrity of the generated data. Using a high-performance aggregation utility, we unified the heterogeneous interaction histories produced by the OpenHands and SWE-agent frameworks into a consistent format. During this phase, we prioritize runtime integrity by systematically discarding corrupted trajectories resulting from unforeseen environment exceptions that prevented the agent from successfully completing its interaction sequence, leaving an incomplete behavioral trace.

\paragraph{Stage 2: Behavioral Pruning and Schema Standardization}
The second stage of our pipeline serves as a rigorous quality-control gate designed to eliminate trajectories that exhibit suboptimal or "hallucinatory" problem-solving patterns. To ensure the model learns from successful, clean interactions, we systematically prune trajectories based on three primary failure modes:

\input{tables/data_stat}

\begin{itemize}[noitemsep, topsep=2pt]
    \item Task Incompleteness: We exclude instances where the agent reached the maximum iteration limit or the trajectory was prematurely terminated by the harness, regardless of whether the resulting patch resolved the issue.
    \item Structural Invalidity: Trajectories are discarded if they yield empty patches or ``cheat'' by altering the test suite (\texttt{test\_patch}) rather than fixing the underlying bug.
    \item Tool-Use Anomalies: We exclude trajectories characterized by malformed interactions, such as illegal concurrent tool invocations or repeated tool usage errors.
\end{itemize}

Following this pruning, the remaining high-fidelity trajectories are standardized into a unified schema. We map framework-specific logs into explicit \texttt{role}, \texttt{content}, and \texttt{tool\_calls} fields while crucially preserving the \texttt{reasoning\_content} from the Minimax-M2.5 model. This ensures the corpus captures the latent chain-of-thought necessary for solving complex software engineering tasks.

\subsection{Security and Integrity: Post-hoc Git Hacking Detection}
To ensure the integrity of the final corpus, we implemented TrajectoryScanner, a validation framework designed to identify and prune trajectories that exhibit "git hacking." This behavior involves agents attempting to exfiltrate repository metadata or bypass legitimate problem-solving by inspecting internal git histories for solution leaks. By leveraging Abstract Syntax Tree (AST) parsing, the scanner audits every generated Bash command, categorizing them into a risk-based hierarchy. We systematically discard any trajectory containing banned operations that expose sensitive metadata (e.g., \texttt{reflog}, \texttt{blame}) or restricted commands (e.g., \texttt{log}, \texttt{diff}) that fail context-specific safety checks. This post-filtering step ensures that the resulting dataset is free from adversarial shortcuts, retaining only those trajectories that demonstrate authentic, safe version control patterns.

\subsection{Dataset Statistics}

We present the comprehensive statistics for \ourdataset across its nine supported languages in \cref{tab:data_stat}. The final corpus comprises 207,489 high-fidelity trajectories, of which 51.7\% incorporate internal chain-of-thought reasoning traces generated by the MiniMax-M2.5 model. In terms of framework distribution, 50.8\% of the trajectories were synthesized via the OpenHands platform. 

The dataset maintains a diverse multilingual balance, with Python (23.2\%), Go (22.6\%), TypeScript (17.8\%), and JavaScript (14.2\%) representing the largest shares of the corpus. Finally, we verified the execution correctness of the generated patches through rigorous unit testing; of the total trajectories, 65,244 successfully resolved the underlying issues, yielding an overall pass rate of approximately 40.6\%.

%% file: tables/data_stat.tex
\begin{table*}[ht]
    \centering
    \begin{tabular}{lccccc}
        \toprule
        \textbf{Language} & \textbf{OpenHands} & \textbf{OpenHands} & \textbf{SWE-agent} & \textbf{SWE-agent} & \textbf{Total} \\
         & \textbf{(w/ thinking)} & \textbf{(w/o thinking)} & \textbf{(w/ thinking)} & \textbf{(w/o thinking)} & \textbf{Traj.} \\
        \midrule
        \textbf{Python} & 11,790 / 4,608 & 12,278 / 4,501 & 13,781 / 4,825 & 10,330 / 4,086 & 48,179 \\
        \textbf{Go} & 12,093 / 4,818 & 13,324 / 4,827 & 10,933 / 4,292 & 10,484 / 4,248 & 46,834 \\
        \textbf{TypeScript} & 8,757 / 3,509 & 9,657 / 3,496 & 10,600 / 3,714 & 7,883 / 3,188 & 36,897 \\
        \textbf{JavaScript} & 6,854 / 2,786 & 7,467 / 2,756 & 8,641 / 3,039 & 6,398 / 2,585 & 29,360 \\
        \textbf{Rust} & 4,412 / 1,975 & 6,283 / 2,385 & 6,345 / 2,388 & 4,695 / 2,024 & 21,735 \\
        \textbf{Java} & 3,175 / 1,270 & 3,463 / 1,288 & 3,767 / 1,366 & 2,771 / 1,152 & 13,176 \\
        \textbf{PHP} & 2,650 / 1,034 & 2,747 / 830 & 2,918 / 1,054 & 2,008 / 830 & 10,323 \\
        \textbf{C} & 159 / 71 & 184 / 72 & 197 / 80 & 143 / 66 & 683 \\
        \textbf{C++} & 58 / 27 & 85 / 33 & 86 / 33 & 73 / 32 & 302 \\
        \midrule
        \textbf{Total} & 55,488 / 20,098 & 49,948 / 20,362 & 57,268 / 20,791 & 44,785 / 18,211 & 207,489 \\
        \bottomrule
    \end{tabular}
    \vspace{-2mm}
    \caption{
    \ourdataset statistics across languages. Each cell denotes Trajectories / PRs. Note that PR counts are non-unique across columns and are not intended for horizontal summation.
    }
    \label{tab:data_stat}
\end{table*}

%% file: sections/3_experiments.tex
\input{tables/main_python}
\input{tables/main_multilingual}

\subsection{Experiment Setup}

\paragraph{Agent Scaffolding.} We employed the multilingual extension of OpenHands \citep{wang2024openhands} and SWE-agent \citep{yang2024swe}, known as MOpenHands (MOH)\footnote{\url{https://github.com/multi-swe-bench/MopenHands}} and MSWE-agent (MSWEA)\footnote{\url{https://github.com/multi-swe-bench/MSWE-agent}} proposed by \citet{zan2026multi}. To adapt for multilingual support and improve reliability, both frameworks updated their prompts and added \texttt{.gitignore} files to exclude disruptive compiled artifacts. MSWE-agent prioritized stability by truncating long observations and fixing command crashes, while MOpenHands resolved critical bugs—notably a tab-to-space rendering error in \texttt{git diff} by utilizing file redirection to ensure patch compatibility in languages like Go.

\paragraph{Agent Distillation.} We perform distillation using the Qwen3-30B-A3B \citep{yang2025qwen3} as base models, specifically the Thinking, Instruct, and Code variants. To evaluate modality impacts, \ourthinkmodel and \ournonthinkmodel are trained on their respective subsets of \ourdataset, while \ourmodel leverages the full corpus. The latter supports dual-mode operation via the \texttt{/think} and \texttt{/no\_think} triggers.\footnote{Technically, this is controlled by the \texttt{enable\_thinking} parameter in the chat template.}
The training process is configured with a learning rate of 1e-5, a batch size of 32, and a warmup ratio of 0.1, supporting a maximum context length of 131,072.

\paragraph{Evaluation Benchmarks and Metrics.} 
We evaluate performance across three key benchmarks: SWE-bench Verified \citep{chowdhury2024introducing}, featuring 500 human-validated Python issues; SWE-bench Multilingual \citep{jimenez2023swe}, which contains 300 tasks distributed across eight languages (C, C++, Go, Java, JS/TS, PHP, Ruby, and Rust); and SWE-bench Pro \citep{deng2025swe}, providing 731 tasks in Python, Go, and JS/TS. While we utilized both agent harnesses for the Verified and Multilingual benchmarks, MOpenHands was excluded from the SWE-bench Pro evaluation due to persistent integration issues; consequently, we report only MSWE-agent results for this benchmark.
Our primary metric is the Resolved Rate (\%), defined as the percentage of issues successfully fixed. During inference, we increased the context window to 262,144 tokens to better handle large codebases.
Inference hyperparameters are detailed in Appendix, and all results represent mean performance across three independent runs.

\subsection{Experiment Results}

\paragraph{Monolingual Evaluation.}
As shown in \cref{tab:swe-bench-results}, the \ourthinkmodel and \ournonthinkmodel variants demonstrate competitive performance on the SWE-bench Verified, achieving an absolute improvement of over 30.0\% compared to their base model counterparts. 
While \ournonthinkmodel lags behind Scale-SWE-Agent by 2.3 points, this gap is likely attributable to the difference in training volume: Scale-SWE-Agent was trained on approximately 71k Python trajectories across 25k unique pull requests (PRs), whereas \ournonthinkmodel utilized only 24k trajectories from 4.5k PRs.
Our primary model, \ourmodel, achieves a 59.3\% and 61.7\% resolved rate under the \texttt{think} and \texttt{no-think} settings, respectively—an 7.7--10.1\% absolute gain over the strong base model’s performance of 51.6\%.

\paragraph{Multilingual Evaluation.}
Evaluation results are presented in \cref{tab:swe-multilingual-results}.
We include Scale-SWE-Agent as the primary baseline for SWE-bench Multilingual (SWE-M); however, we could not obtain meaningful scores for this competitor on the more challenging SWE-bench Pro (SWE-Pro) benchmark, as Scale-SWE-Agent failed on most instances.
Compared to the monolingual results, we observe even larger absolute improvements for \ourmodel. On SWE-M, the resolved rate increased from 33.5\% to 57.1\% using the MSWE-agent framework, while SWE-Pro performance climbs from 28.4\% to 36.8\%.
Notably, our distilled models underperform in the \texttt{think} configuration relative to the \texttt{no-think} variant. We hypothesize that models require more extensive training to successfully develop internal reasoning capabilities rather than simply executing external actions.
Furthermore, the model encounters a distinct framework-specific bottleneck within the MOpenHands harness, where SWE-M performance drops significantly. Further investigation reveals that within MOpenHands, \ourmodel frequently gets trapped in loops in 5–10\% of cases, exhausting the maximum turn limit.


%% file: tables/main_python.tex
\begin{table*}[ht!]
\centering
\resizebox{1.0\linewidth}{!} {%
\begin{tabular}{lllc}
\toprule
\textbf{Model} & \textbf{Base Model} & \textbf{Harness} & \textbf{Resolved (\%)} \\ 

\midrule
\multicolumn{4}{c}{\textbf{Proprietary Models}} \\
\hline
Claude Opus 4.5 \citep{anthropic2025opus45} & - & - & 80.9 \\
GPT-5.2 (xhigh) \citep{openai2025gpt52} & - & - & 80.0 \\
Gemini 3 Pro \citep{google_gemini3pro_2025} & - & - & 76.2 \\

\midrule
\multicolumn{4}{c}{\textbf{Open Source Foundation Models}} \\
\hline
Minimax-M2.5 \citep{minimax2026m25} & - & - & 80.2 \\
GLM-5 \citep{glm5team2026glm5} & - & - & 77.8 \\
Kimi-K2.5 \citep{kimiteam2026kimik25visualagentic} & - & - & 76.8 \\ 
DeepSeek-V3.2 \citep{liu2025deepseek} & - & - & 73.1 \\ 
Qwen3.5-122B-A10B \citep{qwen3.5} & - & - & 72.0 \\ 

\midrule
\multicolumn{4}{c}{\textbf{Open Source Dense Models of Same Size (32B)}} \\
\hline
SWE-Gym-32B \citep{pan2024training}  & Qwen2.5-Coder-32B & OpenHands & 20.6 \\
R2E-Gym-32B \citep{jain2025r2e} & Qwen2.5-Coder-32B & R2E-Gym & 34.4 \\
Skywork-SWE-32B \citep{zeng2025skywork} & Qwen2.5-Coder-32B & OpenHands & 38.0 \\
DeepSWE-32B-Preview \citep{luo2025deepswe} & Qwen3-32B & OpenHands & 42.2 \\
SWE-Mirror-LM-32B \citep{wang2025swe} & Qwen2.5-Coder-32B & MOpenHands & 52.2 \\
SWE-Lego \citep{tao2026swe} & Qwen3-32B & OpenHands & 52.6 \\
SERA-32B \citep{shen2026sera} & Qwen3-32B & SWE-agent & 54.2 \\
daVinci-Dev-32B \citep{zeng2026davinci} & Qwen2.5-32B-Base & SWE-agent & 56.1 \\
SWE-Master-32B-RL \citep{song2026swe} & Qwen2.5-Coder-32B & R2E-Gym & 61.4 \\
SWE-Hero-32B \citep{ludwig2026swe} & Qwen2.5-Coder-32B & OpenHands & 62.2 \\
OpenSWE-32B \citep{fu2026davincienvopensweenvironment} & Qwen2.5-32B-Base & SWE-agent & 62.4 \\
KAT-Dev-32B \citep{zhan2025kat} & Qwen3-32B & - & 62.4 \\

\midrule
\multicolumn{4}{c}{\textbf{Open Source MoE Models of Same Size (30B-A3B)}} \\
\hline
Qwen3-30B-A3B-Inst. \citep{yang2025qwen3} & - & OpenHands & 22.0 \\
Qwen3-30B-A3B-Think. \citep{yang2025qwen3} & - & OpenHands & 22.0 \\
Qwen3-Coder-30B-A3B \citep{yang2025qwen3} & - & OpenHands & 51.6 (50.0$^\ast$) \\
GLM-4.7-Flash \citep{5team2025glm45agenticreasoningcoding} & - & - & 59.2 \\
Scale-SWE-Agent \citep{zhao2026immersion} & Qwen3-30B-A3B-Inst. & OpenHands & 64.0 (59.2$^\ast$) \\

\addlinespace[2pt] \hline \addlinespace[2pt]

\multirow{2}{*}{\ourthinkmodel} & \multirow{1}{*}{Qwen3-30B-A3B-Think.} & MSWEA | MOH & 55.3 | 51.8 \\
& \multirow{1}{*}{Qwen3-Coder-30B-A3B} & MSWEA | MOH & 56.3 | 55.8 \\

\addlinespace[2pt] \hdashline \addlinespace[2pt]

\multirow{2}{*}{\ournonthinkmodel} & \multirow{1}{*}{Qwen3-30B-A3B-Inst.} & MSWEA | MOH & 53.2 | 56.0 \\
& \multirow{1}{*}{Qwen3-Coder-30B-A3B} & MSWEA | MOH & 60.6 | 57.0 \\

\addlinespace[2pt] \hdashline \addlinespace[2pt]

\multirow{1}{*}{\ourmodel (/think)} & \multirow{2}{*}{Qwen3-Coder-30B-A3B} & MSWEA | MOH & 59.3 | 59.1 \\
\multirow{1}{*}{\ourmodel (/no\_think)} & & MSWEA | MOH & {\bf 61.7} | 60.2 \\

\bottomrule
\end{tabular}
}
\vspace{-2mm}
\caption{
Performance comparison on the SWE-bench Verified. OPEN-SWE-AGENT performances are reported as x | y, representing Pass@1 using the MSWE-agent (MSWEA) and MOpenHands (MOH) frameworks, respectively. $^\ast$ indicates score calculated by our evaluation pipeline (with MOpenHands agent harness).
}
\label{tab:swe-bench-results}
\end{table*}

%% file: tables/main_multilingual.tex
\begin{table*}[ht!]
\centering
\resizebox{1.0\linewidth}{!} {%
\begin{tabular}{llcc}
\toprule
\textbf{Model} & \textbf{Base Model} & \textbf{SWE (M)} & \textbf{SWE (Pro)} \\ 

\midrule
\multicolumn{4}{c}{\textbf{Proprietary Models}} \\
\hline

Claude Opus 4.5 \citep{anthropic2025opus45} & - & 76.2 & 52.0 \\
GPT-5.2 (xhigh) \citep{openai2025gpt52} & - & 72.0 & 55.6 \\
Gemini 3 Pro \citep{google_gemini3pro_2025} & - & 65.0 & 43.3 \\

\midrule
\multicolumn{4}{c}{\textbf{Open Source Foundation Models}} \\
\hline

Minimax-M2.5 \citep{minimax2026m25} & - & 80.2 & 55.4 \\
GLM-5 \citep{glm5team2026glm5} & - & 73.3 & 55.1 \\
Kimi-K2.5 \citep{kimiteam2026kimik25visualagentic} & - & 73.0 & 53.8 \\
DeepSeek-V3.2 \citep{liu2025deepseek} & - & 70.2 & - \\ 
Qwen3.5-122B-A10B \citep{qwen3.5} & - & 68.3$^\ast$ & 53.6$^\ast$ \\ 

\midrule
\multicolumn{4}{c}{\textbf{Open Source MoE Models of Same Size (30B-A3B)}} \\
\hline

Qwen3-Coder-30B-A3B \citep{yang2025qwen3} & - & 33.5$^\ast$ & 28.4$^\ast$ \\
Scale-SWE-Agent \citep{zhao2026immersion} & Qwen3-30B-A3B-Inst. & 45.2$^\ast$ & - \\

\addlinespace[2pt] \hline \addlinespace[2pt]

\multirow{2}{*}{\ourthinkmodel} & Qwen3-30B-A3B-Think. & 42.3 | 42.4 & 31.3 | - \\
& Qwen3-Coder-30B-A3B & 44.1 | 45.0 & 32.5 | - \\

\addlinespace[2pt] \hdashline \addlinespace[2pt]

\multirow{2}{*}{\ournonthinkmodel}  & Qwen3-30B-A3B-Inst. & 44.9 | 46.5 & 32.2 | - \\
& Qwen3-Coder-30B-A3B & 51.3 | 44.7 & 34.3 | - \\

\addlinespace[2pt] \hdashline \addlinespace[2pt]

\ourmodel (/think) & \multirow{2}{*}{Qwen3-Coder-30B-A3B} & 47.6 | 43.3 & 33.4 | - \\
\ourmodel (/no\_think) & & {\bf 57.1} | {\bf 48.4} & {\bf 36.8} | - \\

\bottomrule
\end{tabular}
}
\vspace{-2mm}
\caption{
Performance comparison on the SWE-Bench Multilingual (M) and Pro benchmarks. $^\ast$ indicates score calculated by our evaluation pipeline. \ourmodel performance on SWE (M)  are reported as x | y, representing Pass@1 using the MSWE-agent and MOpenHands frameworks, respectively.
}
\label{tab:swe-multilingual-results}
\end{table*}

%% file: sections/4_analysis.tex
\input{tables/analysis}

We aim to address the following research questions.
\begin{enumerate}[noitemsep, topsep=2pt, after=\vspace{2pt}]
    \item {\bf Cross-Harness Generalization}: Does training on trajectories from one agent harness (e.g., MOpenHands) generalize effectively to another (e.g., MSWE-agent)?
    \item {\bf Cross-Lingual Transfer}: To what extent does a Python-only model generalize to multilingual issue resolution tasks?
    \item {\bf Data Filtering Impact}: Does the exclusion of trajectories containing unresolved patches impact downstream agentic performance?
\end{enumerate}

The ablation study results, detailed in \cref{tab:analysis}, provide a critical look at how training configurations influence agentic capabilities across different environments and modalities.

\subsection{Cross-Harness Generalization}
The results indicate that while generalization between evaluation harnesses is achievable, switching agent frameworks incurs a consistent performance penalty. This cross-harness limitation appears to be structural rather than language-dependent, as similar degradation patterns occur across both the Python-centric (SWE-bench V.) and multilingual (SWE-bench M.) benchmarks. The performance drop is driven by a tendency for agents to overfit to the unique interaction patterns, action spaces, or observation formats of their primary training environment. Interestingly, the data reveals a directional preference between the frameworks: models initially trained on MSWEA maintain higher stability and experience smaller performance penalties when transferring to MOH. Conversely, migrating models trained on MOH over to the MSWEA evaluation harness results in much steeper performance drops across both benchmarks, identifying MSWEA as the more robust baseline for cross-harness generalization.


\subsection{Cross-Lingual Transfer}

The transition from a Python-only training set to the "full" (multilingual) \ourdataset demonstrates significant positive transfer, particularly for non-Python tasks.
The most striking improvement is seen in the \texttt{no-think} mode for SWE-bench M., where moving from Python-only to the full dataset boosts the resolved rate from $43.1\%$ to $57.1\%$ (a $+14\%$ absolute gain).
Even on the Python-centric SWE-bench V., adding multilingual data does not dilute performance; instead, it provides a slight boost (e.g., $54.9\% \rightarrow 58.1\%$ in \texttt{think} mode).
While Python-only models generalize reasonably well to other languages, explicit multilingual distillation is essential for peak performance in diverse programming environments, as it allows the model to internalize more generalized logic-flow and syntax-agnostic debugging strategies that circumvent specific language barriers. 

\subsection{Data Filtering Impact}

The results suggest that including trajectories with unresolved patches (the full corpus) is generally superior to training exclusively on "resolved-only" successful trajectories.
Moving from resolved-only to the full dataset yields improvements across almost all categories. On SWE-bench V. (\texttt{think}), we observe a climb from $55.3\%$ to $58.1\%$. The impact is specially evident in multilingual benchmarks; on SWE-bench M., the score improves from $40.5\%$ to $47.6\%$ under the \texttt{think} configuration, and from $49.6\%$ to $57.1\%$ in the \texttt{no-think} variant.
These results suggest that "negative" samples and longer trajectories within the full corpus help the model navigate complex states, even without a successful fix. Excluding unresolved trajectories is counterproductive, as the model benefits from exposure to real-world repository interactions regardless of whether a correct patch was ultimately generated in the training sample.

\paragraph{Key Takeaways.} Our analysis establishes a clear hierarchy of performance drivers: multilingual data and full trajectory inclusion (unresolved cases included) are the primary catalysts for gains. Conversely, harness consistency remains a technical challenge, careful alignment across evaluation environments for peak performance.

%% file: tables/analysis.tex
\begin{table*}[h!]
\centering
\renewcommand{\arraystretch}{1.0}
\begin{tabular}{@{}l cc cc@{}}
\toprule
\multirow{2.5}{*}{\textbf{Experiment / Configuration}} & \multicolumn{2}{c}{\textbf{SWE-bench V.}} & \multicolumn{2}{c}{\textbf{SWE-bench M.}} \\ 

\cmidrule(lr){2-3} \cmidrule(l){4-5}

& \textbf{Thinking} & \textbf{No-Thinking} & \textbf{Thinking} & \textbf{No-Thinking} \\ 

\midrule
\textbf{Cross-Harness Generalization} & & & & \\
\quad MOH $\rightarrow$ MSWEA & 56.0 $\rightarrow$ 53.9 & 56.9 $\rightarrow$ 55.9 & 43.4 $\rightarrow$ 38.8 & 46.9 $\rightarrow$ 43.6 \\
\quad MSWEA $\rightarrow$ MOH & 56.8 $\rightarrow$ 53.9 & 59.5 $\rightarrow$ 55.3 & 44.5 $\rightarrow$ 43.8 & 52.5 $\rightarrow$ 51.1  \\ \midrule
\textbf{Cross-Lingual Transfer} & & & & \\
\quad Python-only $\rightarrow$ Full & 54.9 $\rightarrow$ 58.1 & 58.6 $\rightarrow$ 59.8 & 33.2 $\rightarrow$ 47.6 & 43.1 $\rightarrow$ 57.1 \\ \midrule
\textbf{Data Filtering Impact} & & & & \\
\quad Resolved-only $\rightarrow$ Full & 55.3 $\rightarrow$ 58.1 & 57.7 $\rightarrow$ 59.8 & 40.5 $\rightarrow$ 47.6 & 49.6 $\rightarrow$ 57.1 \\ \bottomrule
\end{tabular}
\vspace{-2mm}
\caption{Evaluation results of distillation using different subset of \ourdataset.}
\label{tab:analysis}
\end{table*}

%% file: sections/5_related_works.tex
\paragraph{\bf SWE Benchmarks.} 
The software engineering evaluation landscape has evolved from foundations like SWE-bench \citep{jimenez2023swe} and SWE-bench-Verified \citep{chowdhury2024introducing} into a sophisticated ecosystem. Modern benchmarks now probe specialized dimensions including multi-modal integration \citep{yang2024swe}, cross-linguistic proficiency \citep{guo2025omnigirl, rashid2025swe, zan2026multi}, and long-horizon reasoning \citep{deng2025swe}. Contemporary evaluations also scrutinize high-level tasks like full-repository generation \citep{ding2025nl2repo}, scientific expertise \citep{duston2025ainsteinbench}, and niche technical competencies \citep{ma2025swe, shetty2026gso}, establishing a rigorous standard for autonomous engineering. Notably, BeyondSWE \citep{beyondswe2026} extends this scope to cross-repository reasoning and dependency-driven migration, further raising the bar for autonomous engineering.

\paragraph{\bf SWE Datasets.} 
Improving LLM programming proficiency requires high-quality, repository-level data. Data scaling strategies generally follow two paths: synthetic generation (e.g., R2E-Gym \citep{jain2025r2e}, SWE-smith \citep{yang2025swe}, and SWE-Mirror \citep{wang2025swe}) and real-world mining. While SWE-Gym \citep{pan2024training} and SWE-rebench \citep{badertdinov2025swe} established thousands of validated instances, Scale-SWE \citep{zhao2026immersion} massively expands this to 100k instances using multi-agent workflows. To optimize training efficiency, SWE-Zero/Hero \citep{ludwig2026swe} utilizes a two-stage distillation pipeline, contributing 300k execution-free and 13k execution-backed trajectories. Similarly, SWE-World \citep{sun2026swe} enables large-scale data utilization by replacing physical environments with learned surrogate models for feedback. Synthesizing these trends, SWE-Lego \citep{tao2026swe} merges authentic PRs with synthetic instances to maximize both precision and training volume.

\paragraph{\bf SWE Models and Agents.} 
Autonomous coding is shifting from external architectures toward model-centric optimization. While early frameworks like SWE-agent \citep{yang2024swe} and OpenHands \citep{wang2024openhands} used environment-based prompting, newer research embeds reasoning into model weights via mid-training \citep{zeng2026davinci}, SFT on expert trajectories \citep{wang2025swe, tao2026swe}, or reinforcement learning \citep{luo2025deepswe, cao2025skyrl}. Parallel to these agents, a modular "agentless" paradigm \citep{xia2024agentless, yang2025kimi} has emerged, streamlining troubleshooting into distinct stages of localization, repair, and verification \citep{he2025sweswiss, xie2025swe}. In this landscape, Orchard \citep{peng2026orchard} bridges the proprietary performance gap by codifying large-scale agentic trajectories into reusable, open-source training recipes.

%% file: sections/6_conclusion.tex
We introduced \ourdataset, the most extensive collection of software engineering agent trajectories to date, featuring 200,000+ traces across nine languages. By addressing the scarcity of large-scale agentic data, this work enables the distillation of complex engineering behaviors into open-source models. Our evaluation shows strong results: a {\bf 61.7}\% resolve rate on SWE-bench Verified, {\bf 57.1}\% on Multilingual, and {\bf 36.8}\% on the rigorous SWE-bench Pro. These findings suggest that combining explicit thinking modalities with high-quality behavioral traces is a highly effective path for developing specialized autonomous agents.

%% file: sections/appendix.tex
\appendix
\clearpage

\twocolumn[{%
 \centering
 \Large\bf Technical Appendices \\ [20pt]
}]

\section{Implementation Details}
The SFT and inference hyperparameters are detailed in \cref{tab:hyperparam} and \cref{tab:infer_hyperparam}.

\input{tables/all_hyp}




%% file: tables/all_hyp.tex

\begin{table}[h]
\centering

\begin{subtable}{1.0\linewidth}
\centering
\resizebox{1.0\linewidth}{!} {%
\begin{tabular}{lc}
\toprule
\textbf{Hyperparameter} & \textbf{Value} \\
\midrule
Learning Rate & 1e-5 \\
Min. Learning Rate & 1e-8 \\
Base Model & Qwen3-30B-A3B \\
Batch Size & 32 \\
Maximum Context Length & 131,072 \\
Warmup Ratio & 0.1 \\
Weight Decay & 0.01 \\
LR Scheduler Type & Cosine \\
Epoch & 3 \\
\bottomrule
\end{tabular}
}
\caption{Distillation hyperparameters.}
\label{tab:hyperparam}
\end{subtable}

\vspace{4mm} 

\begin{subtable}{1.0\linewidth}
\centering
\resizebox{1.0\linewidth}{!} {%
\begin{tabular}{lll}
\toprule
\textbf{Hyperparameter} & \textbf{/think} & \textbf{/no\_think} \\ 
\midrule
Temperature                 & 1.0   & 0.7   \\ 
Top-p                        & 0.95  & 0.8   \\ 
Top-k                        & 40    & 20    \\
Presence Penalty            & n/a   & 1.5   \\
Maximum Turn                & 250   & 250   \\
Maximum Context Length      & 256k  & 256k  \\
\bottomrule
\end{tabular}
}
\caption{Inference hyperparameters.}
\label{tab:infer_hyperparam}
\end{subtable}

\caption{Comprehensive breakdown of \ourmodel hyperparameters.}
\label{tab:main_hyperparams}
\end{table}